# Localization-dependent charge separation efficiency at an organic/inorganic hybrid interface.

Laura Foglia, Lea Bogner, Martin Wolf and Julia Stähler*.

Fritz-Haber-Institut der Max-Planck-Gesellschaft, Dept. of Physical Chemistry, Faradayweg 4-6, 14195 Berlin, Germany

**ABSTRACT:** By combining complementary optical techniques, photoluminescence and time-resolved excited state absorption, we achieve a comprehensive picture of the relaxation processes in the organic/inorganic hybrid system SP6/ZnO. We identify two long-lived excited states of the organic molecules of which only the lowest energy one, localized on the sexiphenyl backbone of the molecule, is found to efficiently charge separate to the ZnO conduction band or radiatively recombine. The other state, most likely localized on the spiro-linked biphenyl, relaxes only by intersystem crossing to a long-lived, probably triplet state, thus acting as a sink of the excitation and limiting the charge separation efficiency.

An efficient light harvesting device relies on high carrier mobilities, low charge injection and ejection barriers and strong light-matter interaction. Inorganic semiconductors, on which current technology is predominantly based, fulfill the first two requirements (1) but their optical bandgap limits the amount of energy that can be converted (2). Strong light-matter coupling, instead, occurs in *organic* semiconductors and chemical design allows for flexible adjustment of absorption and emission spectra (3). Unfortunately, organic-based devices often suffer from low mobilities (4,5) and charge recombination can occur before extraction. The combination of inorganic and organic semiconductors into hybrid structures promises to lead to a new generation of devices that exploit the advantages of both material classes to increase light conversion efficiency (6,7).

Efficient hybrid devices are based on a careful choice of the two materials; the energy level alignment at the interface determines the occurrence of charge and energy transfer processes. Furthermore, the device performance is affected by the relative probability of photoinduced energy relaxation processes that, by competing with the desired energy or charge transfer process at the hybrid interface, lead to an overall loss of harvested energy. In the molecular film, for example, these processes include a) intramolecular vibrational relaxation (IVR), b) internal conversion (IC), c) (non-) radiative recombination to the electronic ground state, d) triplet formation via intersystem crossing (ISC) or e) separation into a charge transfer state, among others. By reducing the exciton lifetime, these mechanisms shorten the exciton diffusion length and lower the probability for charge and energy transfer to occur. Similarly, the diffusion is likely to be less efficient in the case of strongly localized long-lived excited states such as charge transfer and triplet excitons.

The relevance of a given relaxation pathway with respect to the others depends on the relative rate of the processes, which is in turn affected by the energetic separation of the involved energy levels. One prominent example is given by Kasha's spectroscopic rule stating that light emission always occurs from the lowest electronic excited state independent of the excitation density, since for all higher excited states non-radiative decay pathways to the lowest excited state occur on faster timescales than radiative recombination to the electronic ground state (8). The complete understanding of all competing energy relaxation pathways and the identification of the related electronic states on a fundamental level is therefore crucial for the design and optimization of hybrid devices. Since these relaxation mechanisms occur typically on timescales ranging from hundreds of femtoseconds (IC and IVR) to picoseconds (fluorescence) or even microseconds (triplet recombination), their investigation requires the application of time-resolved techniques that are able to access excited state dynamics on such a variety of timescales.

In this paper, we investigate the balance of relaxation pathways in the organic/inorganic hybrid system formed by the spirobifluorene derivative 2,7-bis(biphenyl-4-yl)-2',7'-ditertbutyl-9,9'-spirobifluorene (SP6) and the non-polar (10-10) surface of ZnO and, for comparison, in the dye film deposited on an inert glass substrate. SP6 is constituted by a sexiphenyl backbone and a spiro-linked biphenyl, further decorated by tertbutyl groups. Both, the spiro-linkage and the tertbutyl groups, are used to improve the morphology of the films and their optical properties by the prevention of crystallization (9). SP6 is characterized by a broad absorption spectrum starting at 3.2 eV and peaking at 3.6 eV (10,11), and exhibits very high fluorescence yields, amplified spontaneous emission and lasing (12,13). The wide band gap, large exciton binding energy and metallic surface of ZnO make it a desirable material for transparent electrodes and substrates in optoelectronic applications. Together, they form a type II junction, i.e. the lowest unoccupied molecular orbital (LUMO) is aligned with the ZnO conduction band while the highest occupied molecular orbital (HOMO) lies within the ZnO band gap. Both, charge separation (CS) with electron transfer from the LUMO to the conduction band and Förster type resonant energy transfer from ZnO to SP6, were previously demonstrated by time-resolved photoluminescence (PL) measurements

(10,11). In particular, the CS process was found to be limited by exciton diffusion in the organic layer to the ZnO interface, with an estimated diffusion length of 10 nm at room temperature. Once the exciton has reached the interface, though, charge separation occurs very efficiently within an estimated timescale of 10 ps (11).

Here, the relaxation pathways in SP6 and their influence on CS at the ZnO interface are addressed by the combination of two complementary all-optical techniques: PL and time-resolved excited state transmission (tr-EST) spectroscopy. While the former addresses the radiative recombination from the lowest excited state into the electronic ground state, the latter is sensitive to higher excited state electronic transitions that are activated after photoexcitation, accessing the dynamics of those excited states that are "dark" to PL. Together, they allow to gain a comprehensive picture of the excited state population dynamics in this system and the related energy loss channels. Several relaxation pathways are found to compete with CS. After initial IVR occurring on a few picoseconds timescale, two distinct excited states of comparable lifetime of hundreds of picoseconds, $X_{6P}$ and $X_{2P}$, form. $X_{6P}$ is localized on the sexiphenyl molecular backbone while $X_{2P}$ is most likely a charge transfer exciton with the electron localized in the biphenyl. The two π-systems are decoupled by the spiro-link that seems to hinder internal conversion between $X_{6P}$ and $X_{2P}$. Therefore, two independent relaxation pathways evolving either as a $X_{6P}$ or $X_{2P}$ species are determined either at the photoexcitation stage or directly after, on the ultrafast timescales of IVR. Only $X_{6P}$ is found to radiatively recombine and efficiently charge separate at the hybrid interface. $X_{2P}$, instead, decays exclusively by the formation of a long-lived, most likely triplet state and therewith represents the most important loss channel for the charge separation process. These findings highlight the fundamental influence of exciton localization and excited state (de)coupling to the efficiency of hybrid organic/inorganic systems.

## 1. Methods

The samples are prepared in an ultrahigh vacuum chamber with base pressure below $10^{-10}$ mbar. The ZnO surface is cleaned by repeated cycles of Ar$^+$ sputtering (10 min, $E = 0.75$ keV) and annealing at 800 K for 30 min (14). The glass substrate was sonicated in ethanol and used without further preparation. The molecules are sublimated with a Knudsen-type effusion cell at a base temperature of 570 K, and the substrates are kept at 300 K during deposition. SP6 forms smooth amorphous films on the ZnO surfaces (11), and other substrates were shown to have negligible impact on the ground state vibrational properties (15). The thickness of the films is monitored by a quartz microbalance and is 25 nm on the glass substrate and 20 nm on the ZnO. After deposition, the samples are transferred through air to a cryostat for optical spectroscopy, with a base pressure of $\sim 10^{-6}$ mbar. In all the reported measurements the sample was kept at 100 K using liquid nitrogen.

The excited state dynamics are investigated using time-resolved excited state transmission (tr-EST) in a pump-probe scheme, as illustrated in the inset of figure 1(b). A first laser pulse (pump) optically excites the system at 3.7 eV, i.e. slightly above the SP6 absorption maximum. After a variable time delay, a second laser pulse (probe) is used to monitor the evolution of the pump-induced changes on light transmission. The experiments are performed using a regeneratively-amplified fs (Ti:Sa) laser system (Coherent RegA) working at 200 kHz repetition rate and providing 800 nm light in pulses of 45 fs duration. 50% of the laser power is used to generate the pump: an optical parametric amplifier (OPA) generates 1.85 eV photons, which are frequency doubled to achieve $h\nu_{pump} = 3.7$ eV which is used to excite SP6 close to its maximum absorption. A white light continuum (WLC) ranging from 1.77 to 2.4 eV is generated by focusing the 1.85 eV light into a 3 mm thick sapphire crystal and compressed to 20 fs with a deformable mirror as described in (16). After interaction with the sample, the transmitted white light is filtered by color filters of 10 nm bandwidth in the range between 540 nm (2.3 eV) and 694 nm (1.78 eV) and detected by a photodiode. The pump-induced change in transmission $\Delta T$ is measured with a lock-in amplifier and the values are normalized to the ground state transmission.

The luminescence spectra are generated by excitation of the sample at the same photon energy and detected by a spectrometer, which entrance fiber is placed in front of the optical cryostats window.

To avoid photodamage, the maximum pump fluence is kept to 0.15 nJ per pulse. At the excitation energy, the absorption coefficient of SP6 is $\alpha_{SP6} = 4 \times 10^5$ cm$^{-1}$ (11) and 40% of the incident intensity is absorbed in the films, resulting in an excitation of 0.1% to 0.2% of the molecules.

## 2. Results and discussion

Panel (a) of Figure 1 shows a false color plot of the pump-induced changes in transmission as a function of probe photon energy and time elapsed after photoexcitation, after subtraction of the background at negative time delays. Note the logarithmic x-axis in the range from 1 to 400 ps.

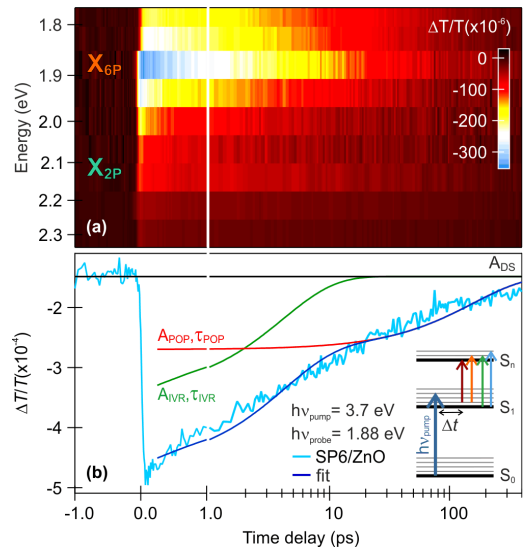

**Figure 1 (a)** False color plot of the pump-induced transmission changes in SP6/ZnO at 100K as a function of pump-probe delay (x-axis) and probe photon energy (y-axis). Note the logarithmic time axis for 1 to 400 ps. **(b)** Tr-EST trace for a representative probe wavelength of 1.88 eV (light blue). An empirical double exponential fit (dark blue) allows for disentanglement of fast (green, 2-6 ps), slow (red, ~250 ps) and long-lived (black, >5 μs) dynamics. **(Inset)** tr-EST scheme. The pump pulse excites the molecule and, after a controlled time delay $\Delta t$, the probe pulse monitors electronic transitions to higher excited states.

The transmission drops abruptly at $\Delta t = 0$ fs, i.e. when both pump and probe pulses overlap temporally on the sample. This indicates the onset of excited state absorption (ESA), which decays for positive delays. The transmission change initially shows a broad energy distribution while, for longer time delays, two resonances are visible. Panel (b) shows the kinetic trace for an exemplary probe photon energy of 1.88 eV (660 nm) before background subtraction. The whole trace shows an amplitude offset (see negative delays) that results from ESA lasting longer than the inverse laser repetition rate, i.e. originating from an excited state with a lifetime >5 μs, as discussed in more detail below. The ESA decay is fitted with an empirical double exponential function, starting at $\Delta t = 250$ fs, where we can assume electronic coherences to be dephased and the time constants to be exclusively related to the evolution of the excited state population distribution (17). On these timescales, the amplitude of the signal is related to the oscillator strength of the probed transition and the electronic population in its initial level. In the empirical fit function

$$\frac{\Delta T}{T} = A_{IVR} \cdot \exp\left[\frac{-(t-t_0)}{\tau_{IVR}}\right] + A_{POP} \cdot \exp\left[\frac{-(t-t_0)}{\tau_{POP}}\right] + A_{DS}$$

$A_{IVR}$ and $A_{POP}$ are the amplitudes of the fast and slow components, respectively, and $\tau_{IVR}$ and $\tau_{POP}$ the corresponding time constants. $A_{DS}$ accounts for the offset at negative delays and is determined independently of the fit.

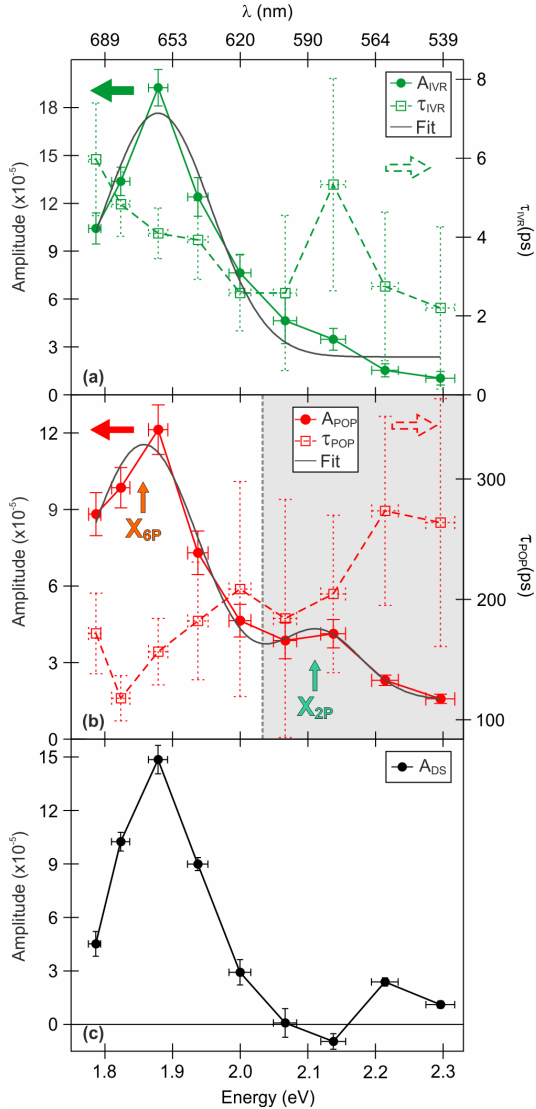

**Figure 2** Fit parameters as a function of probe photon energy. Left axes: amplitudes (full circles and solid line). Right Axes: time constants (empty squares and dashed line). (a) Fast exponential recovery: broad amplitude distribution associated with intramolecular vibrational relaxation; (b) slow exponential recovery: population decay. Two resonances of comparable lifetime are observed; (c) negative delay offset: population of a "dark state".

Qualitatively similar dynamics are observed for all nine different probe photon energies.

Figure 2 shows the parameters resulting from the fit as a function of probe photon energy. The amplitudes are plotted as full circles and solid lines (left axis) and the time constants as open squares and dashed lines (right axis). Panel (a) depicts the amplitude $A_{IVR}$ and time constant $\tau_{IVR}$ characteristic of the fast component. The amplitude clearly peaks around 1.87(5) eV, as determined by a Gaussian fit (gray line) and exposes a broad energy distribution. The time constant shows a strong energy dependence, spanning from 2(2) to 6(2) ps and is shorter for higher resonance energies. These amplitudes and time constants characterize the response of the sample shortly after the pump pulse has populated the excited states by vertical projection of the ground state population. The time constants measured in an optical experiment, where the probe photons are mapping different vertical transitions in the potential energy landscape, is related to the population distribution in the initial state of the transition. Therefore, a broad amplitude distribution and strongly energy dependent lifetime are indicative of a population spread over multiple vibrational levels that relaxes to the bottom of the multiplicity by IVR.

The amplitude $A_{POP}$ and time constant $\tau_{POP}$ of the slower decay are plotted in panel (b). The amplitude distribution is narrower and clearly shows two resonances, labeled $X_{6P}$ and $X_{2P}$ for reasons explained below, at 1.85(3) and 2.12(3) eV respectively. The two resonances have averaged lifetimes of 140(30) ps for $X_{6P}$ and 230(50) ps for $X_{2P}$. Therefore, both lifetimes are on the same order of magnitude as the 300 ps lifetime of the lowest excited state estimated from a diffusion model for the isolated molecule in reference (11). As discussed above, the timescales resulting from a tr-EST experiment depend directly on the population of the initial state of the transition: Two resonances having different time constants must originate from two distinct initial populations, i.e. two separate electronic or vibrational levels.

Additionally, as discussed above, a non-zero offset is observed in the tr-EST at negative delays, with an amplitude $A_{DS}$ that is depicted in panel (c). This absorption is absent without the pump pulse. Its maximum energy position and intensity is comparable to the ones of both the fast and slow decay components. The presence of a pump-induced, non-zero offset at negative delays points at the formation of a very long-lived state. At times before time zero, the probe pulse samples the absorption of an excited state population that has been generated by the previous pump laser pulse, i.e. at a time delay corresponding to at least the inverse of the laser repetition rate. Working at 200 kHz, the excited state lifetime, hence, must exceed 5 μs, thus being four orders of magnitude larger than the singlet lifetime. Therefore, we conclude that $A_{DS}$ is related to the formation of a dark state, e.g. a triplet or a charge transfer exciton (19). The comparable magnitude and spectral dependence of $A_{DS}$, $A_{IVR}$ and $A_{POP}$ also suggests that this is an efficient decay channel and is competitive with radiative recombination and exciton diffusion to the ZnO interface. Consequently, it must occur on a comparable timescale of hundreds of picoseconds.

The determination of the relevance of a specific pathway is complicated by the fact that the decay rates of the multiple relaxation mechanisms sum up to one overall time constant of the population decay. Disentanglement of the contributions becomes possible if the balance of the relaxation processes is altered in a targeted way. Here, this balance is modified by changing the ZnO substrate to glass, where the lowest molecular excited state is no longer resonant with the conduction band but lies within the band gap and interfacial charge separation is quenched. If charge separation is indeed a competitive decay channel, the excited state lifetime is expected to increase when this pathway is closed. Nevertheless, as CS is diffusion-limited in SP6 films (11), only those relaxation pathways that occur on a timescale comparable to exciton diffusion are expected to be affected.

Upon substrate change, we qualitatively observe the same excited state dynamics as on ZnO. The transmission abruptly drops at time zero and recovers biexponentially. The traces also show the non-zero offset at negative delays, indicative of a long-lived state. We apply the same analysis as above and obtain a similar set of fit parameters $A_{IVR}$, $\tau_{IVR}$, $A_{POP}$, $\tau_{POP}$ and $A_{DS}$ as a function of probe photon energy. The amplitudes (not shown) follow the same behavior as for the ZnO substrate and have comparable intensities.

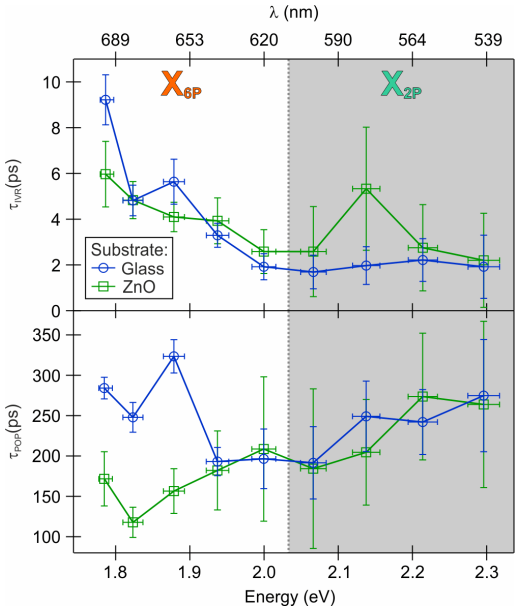

**Figure 3** (a) Comparison of relaxation time constants in SP6 on glass (blue) and on ZnO (green). Top: intramolecular vibrational relaxation. Bottom: population decay. The lifetime of $X_{6P}$ is strongly affected by the ZnO substrate whereas the one of $X_{2P}$ is not.

Figure 3 compares the relaxation time constants of SP6/ZnO (green) with SP6/glass (blue), as a function of probe photon energy. The IVR time constant is plotted in the top panel and the population relaxation below. While (i) the IVR timescale is not affected by the closing of the CS pathway, (ii) the $X_{6P}$ population lifetime increases to 250(30) ps, comparable to the lifetime of $X_{2P}$.

(i) The substrate independence of IVR dynamics implies that this process is not influenced by charge separation to the ZnO. As discussed above, this indicates that IVR occurs on a much shorter timescale than diffusion and, thus, cannot compete with it.

(ii) The influence of the substrate on the $X_{6P}$ lifetime implies that diffusion *can* be a competing process to population decay pathways such as dark state formation and radiative recombi-

nation. The difference of the inverse time constants for the two substrates

$$\frac{1}{\tau_{POP}^{Glass}} - \frac{1}{\tau_{POP}^{ZnO}} = \frac{1}{\tau_{DIFF}}$$

yields an estimate of the *average* diffusion time of 300(100) ps for the 25 nm film.

Remarkably, the charge separation decay channel is open for one of the two observed resonances only: The lifetime of resonance $X_{2P}$ remains *unchanged* for both substrates. As charge transfer is very efficient as soon as the excitons are sufficiently close to the interface, given the large density of states available in the ZnO conduction band, we conclude that the excitons in $X_{6P}$ and $X_{2P}$ have very different diffusion constants. The diffusion length of $X_{2P}$ excitons must be shorter than the average distance to the ZnO interface: the excitons decay before they can charge separate. Finally, no signature of coupling between $X_{6P}$ and $X_{2P}$, e.g. a delayed rise of $X_{6P}$ intensity, is observed in the tr-EST data, suggesting a low internal conversion efficiency.

Excited state transmission alone does not allow to distinguish the nature of the $X_{6P}$ and $X_{2P}$ excitations. The observed ESA resonances could have multiple origins: A) they could be similar SP6 electronic states, located on two kinds of molecules that differ by a chemical modification that occurred during sublimation, B) they could belong to two different spin multiplicities, e.g. singlet and triplet excitons, C) the molecules could aggregate upon deposition and one of the states could be the aggregate population, and D) $X_{6P}$ and $X_{2P}$ could simply be two different states of the SP6 molecules.

We excluded A) by performing proton nuclear magnetic resonance (NMR) experiments, which show identical peak positions for a deuterated benzene solution of SP6 molecules before and after sublimation as well as for the theoretically predicted spectrum (not shown). B) can be ruled out due to lifetime reasons: A triplet exciton is expected to have a significantly longer lifetime than a singlet due to the spin-flip process required for relaxation to the ground state. Instead, both observed states have comparable lifetimes and do not match the much longer lifetime observed for the dark state. Hypothesis C) can be tested by temperature-dependent photoluminescence spectra. In the case of aggregate formation, one photoluminescence peak, the aggregate line, should exhibit markedly different temperature dependence than the monomer emission, since the shape of aggregate emission depends on exciton delocalization, which rapidly decreases with increasing temperature.

Figure 4 shows in green the SP6/glass photoluminescence spectrum after excitation at 3.7 eV, at 100 K (green dots), compared with the spectrum at 10 K (orange triangles). The two spectra are normalized to the maximum emission at 2.96 eV in order to compare the spectral shape. We do not observe any significant difference for the emission shape at different temperatures, i.e. no signature of aggregate formation, from which we can exclude hypothesis C). After ruling out scenarios A)-C), we conclude that D) $X_{6P}$ and $X_{2P}$ are two different states of the SP6 molecule.

Figure 4 also depicts the SP6/ZnO emission spectrum at 100K. Clearly, the emission yield is lower than the emission yield on glass, indicating that the radiative recombination of SP6 is affected by charge separation. Consequently, we associate the radiative state of SP6, i.e. according to Kasha's rule the lowest excited state, to $X_{6P}$. The photoluminescence spectrum exhibits four distinct vibronic peaks, which have been fitted by a sum of Gaussians (violet line and gray dashed curves).

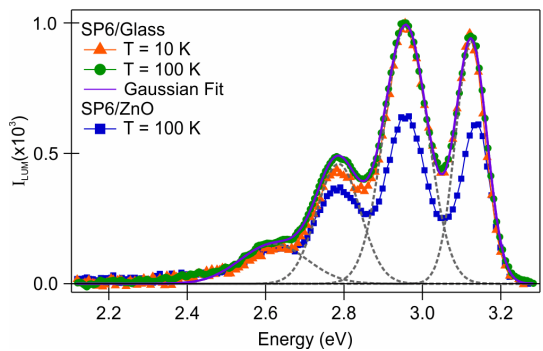

**Figure 4** Photoluminescence spectrum of SP6/glass at 100 K (green dots) compared with the spectrum at 10 K (orange triangles), both normalized at the maximum of the first. The coincidence of the two spectra excludes aggregate formation. The lower yield of the SP6/ZnO emission at 100 K (blue squares) suggests that luminescence is affected by CS. All spectra exhibit four peaks with 160(15) meV separation, corresponding to the CC stretch vibrations.

The obtained peak energy separation of 160(15) meV matches the energy of the CC stretching modes in the SP6 Raman spectra (14). The same energy separation also dominates the emission spectra of the majority of phenylenes, and has been assigned to the symmetric inter-ring CC stretch at 159 meV (20). The peak positions and FWHM resulting from the fit are listed in table 1, where they are compared with those of sexiphenyl (6P) reported in reference (21).

**Table 1. Photoluminescence transition energies**

|  | 0-0 line | 0-1 line | 0-2 line | 0-3 line | Species |
|---|---|---|---|---|---|
| Energy (eV) | 3.13(1) | 2.96(1) | 2.779(5) | 2.612(5) | SP6 |
| FWHM (meV) | 94(7) | 131(6) | 131(5) | 200(5) | |
| Energy (eV) | 3.14 | 2.97 | 2.79 | 2.62 | 6P Ref. (21) |
| FWHM (meV) | 75.00 | 112.00 | 120.00 | — | |

The emission photon energies and separations coincide within the experimental error, while the full width at half maximum is systematically larger in SP6. This difference is attributed to the lack of crystallinity in the SP6 film, opposite to the well-ordered needle like structure of condensed 6P, increasing the disorder in the film and thus the width of the emission lines. From this comparison it is evident that the emission spectra of SP6 and 6P are nearly identical, despite the substantial differences in the molecular structure. This coincidence strongly suggests that the emitting excited state, $X_{6P}$, is localized in the sexiphenyl backbone and, more importantly, is *not* affected by the presence of the second π-system in the biphenyl. Since the two π-systems are separated by the non-conjugated spiro-link, it is likely that $X_{2P}$ is localized on the biphenyl. Excited state localization in distinct molecular sections is observed in similar systems where conjugated segments are divided by non-conjugated bridges, e.g. spiro-linked polyfluorenes (22) and stilbene, naphthylene and anthrylene derivatives (23). This would imply that $X_{2P}$ is an intramolecular charge transfer exciton. Its long lifetime, being comparable to the one of $X_{6P}$, is most probably ascribable to the decoupling of the two orthogonal π-systems caused by the spiro-link, which prevents efficient internal conversion. Thus, $X_{2P}$ excitons are found to decay exclusively via ISC within 230 ps and therewith represent the most prominent loss channel in this system.

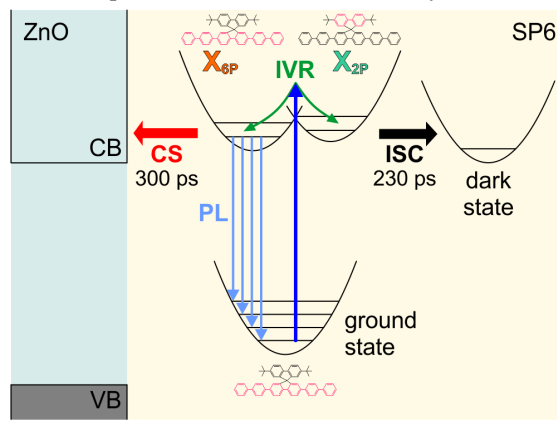

**Figure 5** Relaxation pathways at the SP6/ZnO interface. After photoexcitation the molecules undergo IVR to two distinct excited states $X_{6P}$ and $X_{2P}$ localized on the two π-systems of SP6, respectively. The former decays radiatively (PL), via dark state (DS) formation and via charge separation to the ZnO. The latter decays exclusively via dark state formation and is the most prominent loss channel in the system.

## 3. Conclusions

The combination of two complementary spectroscopic techniques allows to gain a comprehensive picture of the relaxation pathways in the organic/inorganic hybrid film SP6/ZnO, as depicted in figure 5. Photoexcitation brings the system in a highly non-equilibrium state. Thereafter, two excited states of comparable lifetime, $X_{6P}$ and $X_{2P}$, are populated involving intramolecular vibrational relaxation within a few picoseconds. Whether this population process occurs via a higher lying additional excited state or whether both $X_{6P}$ and $X_{2P}$ are resonantly excited cannot be disentangled here; neither does the broad onset of absorption peaking at 3.6 eV (11) provide further insight into the excitation mechanism. Excitation energy-dependent studies of the population *dynamics* could possibly elucidate this question. At later times, however, the $X_{6P}$ excitons, which are localized on the sexiphenyl backbone, decay efficiently by radiative recombination and charge separation to the ZnO conduction band within 140 ps. Instead, the $X_{2P}$ excitons, probably localized in the spiro-linked biphenyl, decay exclusively by dark state formation and represent the main energy loss channel for the charge separation process at the interface. The localization of $X_{2P}$ is attributed to the decoupling of the π-systems by the spiro-link, which apparently hinders excited state internal conversion. It is noteworthy that the weak coupling of $X_{6P}$ and $X_{2P}$ leads to two independent relaxation pathways of the optical excitation in SP6. This means that the "fate" of each exciton is decided on during or directly after photoexcitation, when it either evolves as a diffusive and radiative $X_{6P}$ or as a non-charge-separating, non-radiative $X_{2P}$ species. These findings emphasize the fundamental influence of exciton localization and excited state decoupling on the evolution of excited state population in an organic semiconductor and, consequently, on the efficiency of charge separation processes that are paramount for the functionality of hybrid organic/inorganic systems.

**\*Corresponding Author**

Address:
Physical chemistry department, Fritz-Haber-Institut der Max-Planck-Gesellschaft, Faradayweg 4-6, 14195 Berlin, Germany
Email address:
staehler@fhi-berlin.mpg.de

**Acknowledgments**

We acknowledge fruitful discussions with K. Campen, S. Blumstengel, F. Henneberger, and R. Ernstorfer and thank C. Richter for sample preparation. This project was partially funded by the Deutsche Forschungsgemeinschaft through Sfb 951 and the European Union through grant No. 280879-2 CRONOS CP-FP7.